\documentstyle[aas2pp4,epsfig]{article}
\received{}
\accepted{}
\journalid{}{}

\slugcomment{}
\lefthead{Korchagin, Vorobyov}
\righthead{}
\newcommand{\be}{\begin{equation}}
\newcommand{\ee}{\end{equation}}
\begin{document}

\title{Chemical Abundance Gradients in the Star-Forming Ring Galaxies}
\author{Vladimir Korchagin$^1$}
\affil{Institute of Physics, Stachki 194, Rostov-on-Don, Russia\\
Email: vik@rsuss1.rnd.runnet.ru}
\altaffiltext{1}{National Astronomical Observatory, Mitaka, Tokyo 181, Japan}
\author{Eduard Vorobyov}
\affil{Institute of Physics, Stachki 194, Rostov-on-Don, Russia\\
Email: edik@rsuss1.rnd.runnet.ru}
\and
\author{Y.D. Mayya}
\affil{Instituto Nacional de Astrofisica, Optica y Electronica, Apdo
Postal 51 y 216, C.P. 72000, Puebla, M\'exico \\
Email: ydm@inaoep.mx}
\vskip 0.5cm
\affil{\it Accepted --- April 1999.
To appear in Astrophysical Journal}
\begin{abstract}
Ring waves of star formation, propagating outwardly in the galactic disks,
leave chemical abundance gradients in their wakes. We show that the relative
 $[Fe/O]$ abundance gradients in ring galaxies can be used as a tool
for determining the role of the SNIa explosions in their chemical enrichment.
We consider two mechanisms which can create outwardly propagating star 
forming rings in a purely gaseous disk --- a self-induced wave and a density 
wave, and demonstrate that the radial distribution of the relative $[Fe/O]$ 
abundance gradients does not depend on the particular mechanism of the wave 
formation or on the parameters of the star-forming process. We show that 
the  $[Fe/O]$ profile is determined
by the velocity of the wave, initial mass function, and the initial chemical
composition of the star-forming gas. If the role of SNIa explosions is 
negligible in the chemical enrichment, the ratio $[Fe/O]$ remains constant 
throughout the galactic disk with a steep gradient at the wave front. If SNIa 
stars are important in the production of cosmic iron, the $[Fe/O]$ ratio has 
gradient in the wake of the star-forming wave with the value depending 
on the frequency of SNIa explosions.
   
\end{abstract}
%
%
      \section{INTRODUCTION}
%

Theoretical modeling of chemical abundance gradients in galaxies provides
an important tool for understanding galactic evolution. However, chemical 
evolutionary models applied for the dynamics of galaxies on a Hubble time 
scale encounter some principal difficulties  (Shore \& Ferrini 1995).
The chemical evolution of galaxies is usually considered in the one-zone 
approximation with a postulated accretion rate and some assumed law of 
star formation. These models do not take into account global hydrodynamic
processes of the redistribution of the matter in the galactic disks.
The spiral density waves, viscosity, and the disk-halo-disk circulation
work in different ways to redistribute matter within the galactic 
disks during the galactic evolution, which considerably complicates the problem.

We discuss one class of galaxies where the modern theories of the chemical 
evolution can be successfully applied. These are the starburst galaxies 
with the large scale rings of an enhanced star formation. These galaxies 
are believed to be the result of recent galactic collisions, and result 
from the passage of a compact companion galaxy through the disk of a galaxy 
along its rotation axis (Lynds \& Toomre 1976). Such  collisions produce 
the wave of enhanced star formation, and are responsible for the nature of 
the large scale rings of the young massive stars observed in some galaxies.

The time-scale of bursts of star formation in the galaxies similar to the 
Cartwheel is much shorter compared to the Hubble time, making secular 
hydrodynamic processes unimportant. The chemical evolution in the fast 
waves of star formation is governed therefore by a few parameters which 
can be measured, or at least estimated, and ring galaxies provide a 
possibility to determine the relative role of  SNIa and SNII in the 
enrichment of the interstellar matter by heavy elements.

It is known that there are two physically different sources of heavy 
element production. The $\alpha$-elements  and oxygen
are mainly produced in massive stars experiencing SNII explosions 
(Woosley \& Weaver 1995), and iron peak elements 
are substantially contributed by SNIa which are assumed 
to be the end product of close binary evolution (Tsujimoto et al. 1995).
The relative role of these mechanisms is somewhat unclear due to 
uncertainties in knowledge of the input parameters, and uncertainties of 
stellar evolutionary models (Ishimaru \& Arimoto 1997; Gibson, Loewenstein \& 
Mushotsky 1998).
 
The clock of the SNIa explosions and thus the enrichment due to SNIa is 
substantially delayed with respect to SNII. Such a delay in iron chemical 
enrichment compared to the $\alpha$-elements will reveal itself in the 
star-forming ring galaxies as a relative abundance $[Fe/\alpha]$ radial 
gradient. The value of this gradient depends only on a few parameters
such as the velocity of the wave, the slope of the initial mass function, 
and on the fraction of close binary systems producing iron via type SNIa 
explosions. 
The number of ``free'' parameters is hence drastically reduced, and 
ring galaxies give us unique opportunity to make observational conclusions 
about the role of SNIa explosions in chemical evolution.

%
	\section{CHEMICAL EVOLUTION MODEL}
%

As in our previous paper (Korchagin et al. 1998), we discuss two possible
mechanisms of the formation of large-scale star-forming rings. 
The conventional scenario assumes that rings originate in the direct 
collisions of disk galaxies with their companions. 
Instead, we consider the possibility that galactic collision only 
plays the role of a ``detonator'' stimulating further self-propagating 
star formation. 
This model assumes that the wave is a self-organized phenomenon,
and is similar to the ``fire in the forest'' models discussed earlier.

The most observationally studied ring galaxy is the Cartwheel,
and we adopt therefore parameters of the model obtained for this
archetype galaxy.
The velocity of the wave was chosen equal to 90 km/s --- the value, which 
gives the best fit to the optical surface brightness gradients
observed in the Cartwheel galaxy (Korchagin et al. 1998).
In both models we use the Salpeter
IMF ($\alpha = 1.35$) with masses of stars located in the interval
0.5~M$_\odot \leq m_s \leq 50$~M$_\odot$.
It is remarkable however, that the ratio of iron to oxygen
production in the wave does not depend on the parameters determining
the rate or the efficiency of star formation. This ratio is mainly determined
by the IMF slope, and by the heavy element output of the SN
explosions. This fact makes the theoretical predictions of the radial
dependence $[Fe/O]$ ratio quite robust.

The mathematical formulation of both models, as well as 
the parameters prescribing the rate of star formation
can be found in Korchagin et al. (1998). 

The basic ideas of the chemical evolution model  which we 
incorporate here to describe abundance gradients in the
starburst ring galaxies were outlined by Tenorio-Tagle (1996).
The process of element enrichment in galaxies can be summarized
as follows. The freshly formed elements ejected by the SN
explosions fill the hot interiors of the bubbles, 
which were created earlier by the stellar winds, and/or
by the release of energy in a solitary, or multiple SN explosions. 
The new elements are mixed with pre-existing matter 
through evaporation of a part of the cold shell surrounding the bubble,
before they are expelled into the halo of the galaxy in a hot,
SN-driven winds (``chimneys''). Hot enriched gas eventually
cools and ``rains'' back on the galactic disk completing the
circle of the element enrichment in a new event of star formation.

Recent observations support this picture. Kobulnicky (1999) did not find 
any evidence for metal abundance enhancement in the vicinity of young 
star clusters, concluding that the freshly ejected materials are stored 
in a hot, $10^6$ gas phase. 
Elmegreen (1997), considering processes of mixing and
contamination of interstellar gas after passage of a spiral density wave,
came to a similar conclusion. He finds that
there is no element mixing on time scales  of a few $\times 10^7$ years.

The following processes determine spatial distribution of heavy elements 
in the wake of a star-forming  wave.

{\it a) Stellar Evolution:} The new-born stars ``inherit'' heavy element 
abundances of the pre-existing gaseous disk, and hence their atmospheres
are not expected to contain the heavy elements produced in the current 
starburst. Gas released by the low-mass single stars during their red 
giant phase and the SN explosions with mass $m<11$ is assumed to have 
oxygen and iron abundances equal to that at the birth epoch of the stars.
We take this abundance as to be one fifth of the solar. 
The fraction of mass returned to the
interstellar medium by stars with $m<11$ during their evolution is taken
from  K\"oppen \& Arimoto (1991).

Stars with masses $m > 11 M_{\odot}$ experience SNII explosions
and release freshly formed elements. We use in our simulations the
oxygen and iron yields taken from
models of Woosley \& Weaver (1995) and Nomoto et al. (1997).

The SNIa result from C-deflagration of white dwarfs in binary
systems with masses of progenitors $\le 8 M_{\odot}$.
The nature of the SNI progenitors is rather unclear, and different scenarios
give the lifetimes of SNIa progenitors in the interval $\sim 0.1 - 0.6$ Gyr 
(Branch 1998).
In our simulations, we consider therefore three different  mass intervals
for the SNIa progenitors. 
Namely, we assume that SNIa are produced in binaries with primary
stars of 2.5--8 $M_{\odot}$, 2.5--6 $M_{\odot}$, and 2.5--3.5 $M_{\odot}$. 
Hence the first supernovae explode at $4\times10^7$, $7\times10^7$ and
$2.66\times10^8$ years after the start of the burst, for the three mass
intervals. In the latter case, SNIa become unimportant in the heavy element 
enrichment by fast ring waves of star formation such as inferred in Cartwheel.
We adopt the nucleosynthesis output for SNIa from the updated
W7-model of Thielemann,  Nomoto \& Hashimoto (1993) with  fraction of binaries 
$f_{SNI}$ equal to 0.04 --- the value which Kobayashi et al. (1998) found to be
adjusted to reproduce the chemical evolution in the solar neighborhood.

{\it b) Evaporation:}  
There are two sources of mixing of the chemically unprocessed gas 
with the freshly formed elements inside the hot bubble.
The first mechanism is the classical thermal conductivity, which results
typically a few thousand solar masses being evaporated from the shell
surrounding the hot bubble
(e.g. Shull \& Saken 1995). A larger potential source of mixing,
but far more uncertain, is the photo-evaporation or shock
ablation of the ambient clumpy medium. We consider therefore the following
models of the chemical enrichment in the Cartwheel.
In the first model, we assume that
the freshly formed elements are instantaneously mixed with the remaining
ambient gas, or equally, that all remaining ambient gas is transformed 
into a hot phase by the star formation process. In the second model, 
we consider the conductive evaporation of shells to be the source of 
mass in the bubble. For comparison, we also computed the $[Fe/O]$ gradients
for pure SN ejecta without mixing.

Using expression for the conductive mass-loss of the shell in the adiabatic
bubble (Shull \& Saken 1995) and assuming $10^{51}$ erg of energy released by
each SN and the shell expansion time of $10^6$ yrs, one can write the 
expression
for the SN-evaporated mass in units  $10^7 M_{\odot}$, $10^6$ yr and 1 kpc as: 

\be
   M_{evap}  = (0.7\times10^{-4})
   n_0(r,t)^{-2/35} \kappa^{2/7}
\ee
Here,  $n_0(r,t)$ is the density of the ambient gas, and $\kappa < 1$ 
is a scaling factor.   

{\it c) Cooling:} The hot processed gas streams out in the halo,
and the subsequent abundance gradients are determined by the
concurrent processes of cooling and diffusion. Using Raymond's et al. (1976)
cooling coefficient we find that the cooling time of hot X-ray emitting
gas with the temperature $\sim$ few $\times 10^6 K$ and the density
$10^{-2}-10^{-3}$~cm$^{-3}$ is about a few $\times (10^7 - 10^8)$ yrs which
is large, or comparable to the dynamical time of the starburst.
At the temperature interval typical for the X-ray emitting halo gas, 
the heavy elements, particularly iron, dominate
the cooling process, accounting from half to three-forth of the total
cooling rate (Raymond et al 1976). The heavy element abundances in the
Cartwheel are deficient by factor of ten as compared to the Orion Nebula
in our galaxy (Fosbury \& Hawarden 1977), which will increase the
estimated cooling time. We assumed therefore
that cooling is unimportant during the starburst time in the Cartwheel.

{\it d) Diffusion:} The heavy ion admixtures diffuse into a hot hydrogen
plasma of halo, decreasing possible abundance gradients produced by the
SN explosions. The diffusion of light ions is effective on the
time scale of the starburst (Tenorio-Tagle 1996), but is ineffective
for heavy, highly ionized ions.  In plasma with temperature of a 
few $\times 10^6$ K, the oxygen will be ionized up to OVI, and iron up to 
FeXI--FeIV. The mean free paths of oxygen and iron in plasma with 
temperature $5\times 10^6$ K and density $10^{-3}$ will be about 50pc 
and 12pc correspondingly, which gives values of the dimensionless diffusion 
coefficients $D_O \approx 5\times 10^{-3}$ and $D_{Fe} \approx 6\times 10^{-4}$.
Our simulations show, that with such values, the diffusion of heavy
elements is unimportant on the time-scales of passage wave front in the disk. 

{\it e) Grain Formation:} For the pressure typical in a SN-processed
gas, the grain formation is effective for temperatures below 1000 K 
(Gail \& Sedlmayr 1985). Hot halo gas will preserve therefore the abundance
compositions which were formed during the ejecta. 
However, if the hot gas had enough time to cool, iron will be selectively 
depleted with respect to oxygen due to grain formation.
 
Assuming that the SN ejecta mix with all the remaining ambient gas, 
we can write the equation describing the spatial distribution of the heavy 
elements in the wake of the star-forming wave:
\begin{eqnarray*}
{\partial M_i(r,t) \over \partial t } =& -Z_i^{0} B(r,t)  
+ D_i {\partial^2 M_i \over \partial r^2} \\
                                      +&
\int_{M_{min}}^{M_{max}} B(r,t-\tau_{m}) \phi (m) R_i(m)dm 
\quad (2)
\end{eqnarray*}
Here $B(r,t)$ denotes the birth rate of stars as a function of time and 
radius of the galactic disk, $M_i$ is the surface density of an i-th element, 
$Z_i^{0}=M_i^0/M_{GAS}$ is the  initial abundance of the i-th element in the 
star-forming gas $\phi (m)$ is 
the initial mass function with minimum of maximum masses $M_{min}$ and
$M_{max}$ respectively. $D_i$ is the diffusion coefficient  
of i-th element, and $M_{GAS} $ is the initial constant surface 
density of gas. The coefficient $R_i(m)$ gives ratio of mass of the 
ejected element to the mass of a star $m$. 

In a similar way, we can write the equation governing the chemical evolution
behind the wave
when the stellar ejecta are mixed with gas thermally evaporated from 
the shells with its mass prescribed by the equation (1). 

Once the IMF is fixed, the birth and death rates can be determined at each 
radial grid zone 
making use of the lifetimes of stars. Then the 
heavy element distributions can be computed by means of equations (1) and (2).

\section{Results }

Figure 1 illustrates the $[Fe/O]$ enrichment produced by the wave of    
star formation for the three different scenarios of the element mixing and
three different mass intervals of the pre-SNIa stars. The position of the wave
taken at time $t=240 Myr$ corresponding to the present location of the 
outer ring 
in the Cartwheel galaxy. The left frames show the abundance gradients for the
Nomoto et al. (1997) SNII outputs, and the right frames present the results
of simulations for the Woosley \& Weaver (1995) SNII iron and oxygen outputs.

The bottom frames of Figure 1 show the $[Fe/O]$ profiles if masses 
of pre-SNIa stars 2.5--3.5 $M_{\odot}$. In this case, all SNIa progenitors have 
lifetimes longer than the time of the wave propagation , and the role of SNIa  
stars in the enrichment  is negligible. The solid lines of the Figure 1 
show the $[Fe/O]$ radial dependence when hot SN ejecta are mixed with the
part of ambient gas evaporated from the SN shells. For comparison,
we also plot the abundance gradients when the freshly synthesized elements are
instantaneously mixed with the all ambient gas (dashed lines), and the 
abundance gradients produced by the pure SN ejecta (dotted lines). 

Figure 1 depicts more sharp growth of the oxygen abundance as compared 
to the iron one in front of the wave of star formation. 
Such behavior directly follows from the 
yields of iron and oxygen by the SNII explosions. In the model of Woosley \& 
Weaver (1995) the oxygen yield increases with stellar mass until the progenitor 
mass reaches approximately 40 $M_{\odot}$, and then tends to saturate. 
Supernova models of Nomoto et al. (1997) give rapid increase of oxygen yield 
with the mass of progenitor. The iron yield on the contrary decreases with 
stellar mass in both the models. Therefore, SNII with moderate masses of 
progenitor are responsible mainly for the iron production, whereas SNII 
with progenitors of higher mass produce mainly oxygen. 

Large $[Fe/O]$ gradients in the region of the wave of star formation are 
changed to slower growth of the $[Fe/O]$ ratio towards the center of galaxy 
when metal enrichment due to SNIa events start to become important.
The top and intermediate frames of Figure 1 show the radial profiles 
of $[Fe/O]$ for the SNIa progenitors with masses 2.5--8 $M_{\odot}$ and 
2.5--6 $M_{\odot}$, respectively. The ``pollution'' by the SNIa hot ejecta 
causes the increase of the $[Fe/O]$ gradient towards the center of the disk.
The effect is obviously most noticeable for the pure SN ejecta (dotted lines).
The element mixing diminishes the effect, which however remains
about 0.05 dex in the Fe/O ratio in both SNII output models and both
models of mixing of newly-formed elements.
 
The radial abundance profiles of oxygen and iron obtained for the density 
wave are much similar to those obtained for the induced wave of star formation. 
The radial distributions of relative abundance $[Fe/O ]$ are qualitatively 
independent of the particular model of star formation and  
 the uncertain parameters such as the rate of 
star formation and the ``efficiency'' of star formation. 
This makes our conclusions robust, and the relative 
abundance gradients $[Fe/\alpha ]$ might be used as a tool for 
determining the relative role of SNIa in the chemical enrichment.

\section{DISCUSSION}

The nebular abundances measured by
Fosbury \& Hawarden (1977) in the Cartwheel
pertain to the heavy element abundances  before the passage of the wave.
Therefore, observations
in the inner ring regions are required to test our predictions. 

Interstellar abundances can be inferred however using X-ray emission 
from hot halo gas. 
The launch of the Advanced X-ray Astrophysics Facility ($\it AXAF$)
with its spatial resolution of $\approx 0.^{\prime \prime}5$ would
allow to measure the heavy element gradients in the ring galaxies
of star formation.

If the hot gas has enough time to cool after the passage of the wave
of star formation, and the newly-formed elements reside in the warm
ionized or the cold phases of interstellar medium, then abundances 
can be inferred using metallic absorption lines
towards background objects. The absorption lines of most atoms found in the
interstellar medium occur at ultraviolet wavelengths, and hence
abundance determinations require satellite observations.
With the launch of {\it Hubble Space Telescope} interstellar abundances
are now available for a number of atoms, including iron and the $\alpha$
elements $O, N, C, Mg$ etc. (Savage \& Sembach 1996). 

In external galaxies, the measurement of interstellar abundances depends
on the availability of background sources. The most widely
used background sources are quasars, using which abundances in halos of
galaxies and in the intervening clouds have been determined (e.g. Morton
et al. 1980). In a recent study, Gonz\'alez-Delgado et al. (1998) have
detected interstellar absorption lines towards the bright starburst
nucleus of NGC\,7714 in the ultraviolet spectrum taken with the {\it
Hubble Space Telescope}. Hence, when background sources are available the
present technology allows for the measurement of interstellar absorption lines.
In ring galaxies, the measurement of interstellar abundance thus depends
on the inclination of the ring to the line of sight. In cases, where the
ring is favorably aligned, the spectrum of center regions and the ring HII
regions are likely to contain interstellar absorption lines.


\begin{acknowledgments}
VK acknowledges Prof. S. Miyama for hospitality, and National Astronomical
Observatory of Tokyo for providing COE fellowship.

\end{acknowledgments}
\newpage
%

\newpage

\figcaption{The $[Fe/O]=log({M_{Fe} \over M_{O}})-log{({M_{Fe} 
\over M_{O}})}_{\odot}$ relative abundance profiles  for the wave propagating
in the gaseous disk with $Z_\odot/5$ element abundance. 
The left frames give the results of simulations for the
Nomoto et al. (1997) SNII outputs, and the three right frames show the
results of simulations for the Woosley \& Weaver (1995) SNII outputs.
Top and middle frames --- SNIa progenitors contain relatively
massive stars (2.5--8 and 2.5--6 $M_{\odot}$). The SNIa explosions substantially
contribute to the iron enrichment behind the wave. Bottom frames --- the SNIa
progenitors contain low-mass stars (2.5--3.5 $M_{\odot}$), and SNIa do not
participate in the heavy element contamination. 
Different line types
correspond to the 3 mixing schemes we have adopted --- 
the realistic case of only a part of the ambient gas mixing with 
newly-formed elements (solid lines), 
mixing with all the ambient gas instantaneously (dashed lines), 
and no mixing with the ambient gas (dotted lines).
} 

\begin{figure}[htb]
\vspace*{-3.5cm}
\hspace*{5.0cm}
\centerline{\epsfig{figure=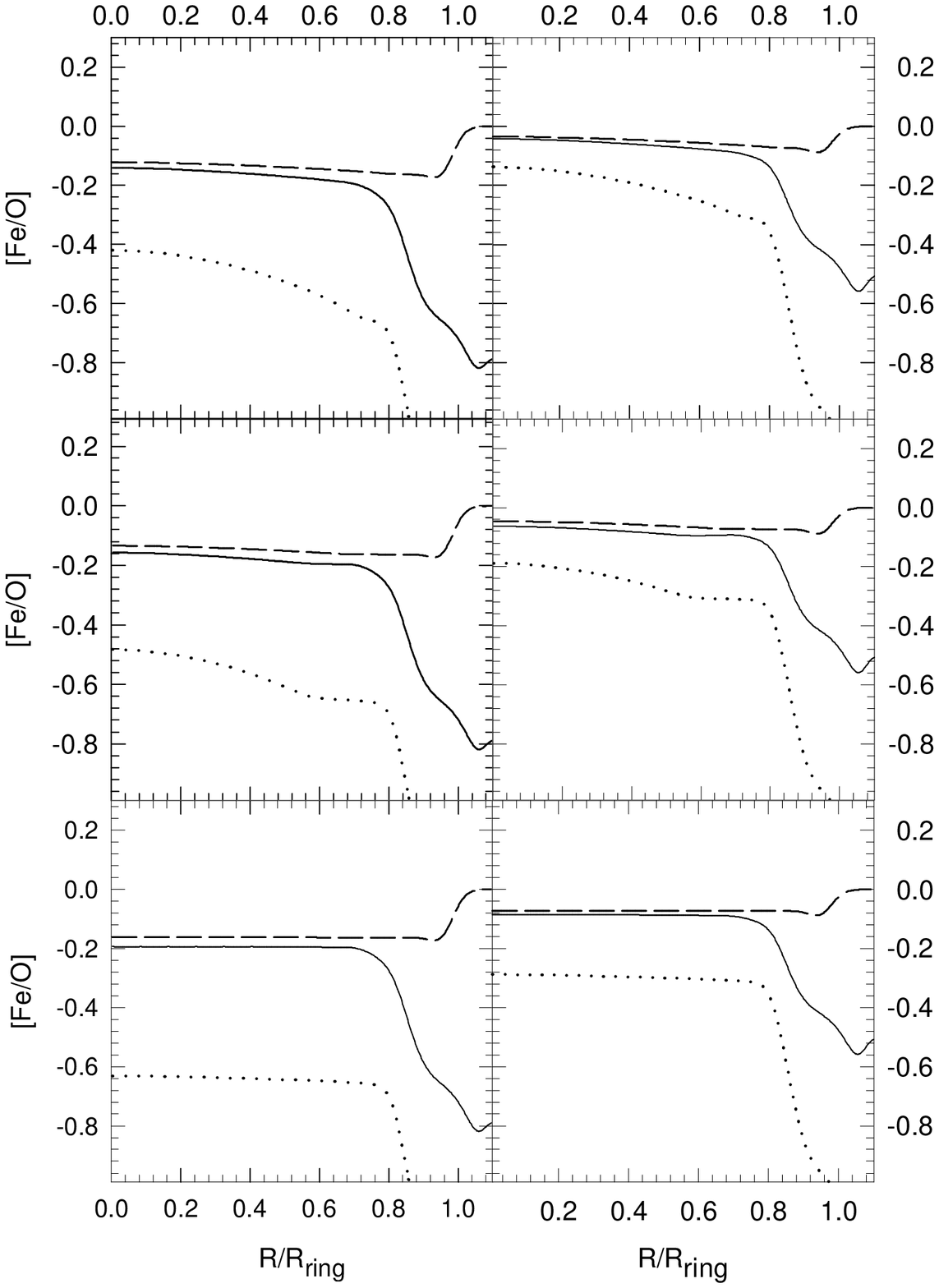}}
\end{figure}


\begin{thebibliography}{}

\bibitem{00}
Branch, D. 1998, ARA\&A, 36, 17

\bibitem{2} 
Elmegreen, B.G. 1997, astro-ph/9712354

\bibitem{1a} 
Fosbury, R.A.E., \& Hawarden, T.G. 1977, MNRAS, 178, 473

\bibitem{3a}
Gail, H.-P., \& Sedlmayr, E. 1986, AA, 166, 225

\bibitem{4} 
Gibson, B.K., Loewenstein, M., \& Mushotsky, R.F. 1997, MNRAS, 290, 623 

\bibitem{4a} 
Gonz\'alez-Delgado, R.M. et al. 1998, astro-ph/9810331

\bibitem{5} 
Ishimaru, Y \& Arimoto, N. 1997, PASJ, 49, 1

\bibitem{11}
Kobayashi, C., Tsujimoto, T., Nomoto, K., Hachisu, I., \& Kato, M. 
      1998, ApJL, 503, 155

\bibitem{9}
Kobulnicky, H.A. 1999, astro-ph/9901260

\bibitem{10}
K\"oppen, J, \& Arimoto, N. 1991, A\&AS, 87, 109

\bibitem{7}
Korchagin, V., Mayya, Y.D., \& Vorobyov, E.I., Kembhavi, A.K. 1998, 
ApJ, 495, 757

\bibitem{13}
Lynds, R., \& Toomre, A. 1976, ApJ, 209, 382

\bibitem{13b}
Morton, D.C. et al. 1980, MNRAS, 193, 399

\bibitem{13c} Nomoto, K., Hashimoto, M., Tsujumoto, T., Thielemann, F.-K.,
Kishimoto, N., Kubo, Y., \& Nakasato, N. 1997, Nucl. Phys., A616, 79c

\bibitem{13d} Raymond, J.C., Cox, P.D., \& Smith, B.W. 1976, ApJ, 204, 290

\bibitem{14}
Savage, B.D., \& Sembach K.R. 1996 ARA\&A, 34, 279

\bibitem{15}
Shore, S.N., \& Ferrini, F. 1995, Fundam. Cosm. Phys., 16, 1

\bibitem{15a} Shull, J.M., \& Saken, J.M. 1995, ApJ, 444, 663

\bibitem{16a}
Tenorio-Tagle, G. 1996, AJ, 111, 1641

\bibitem{17}
Tsujimoto, T., Nomoto, K., Yoshii, Y., Hashimoto, M., Yanagida, S.
   \& Thielemann, F.-K. 1995, MNRAS, 277, 945

\bibitem{17a}
 Thielemann, F.-K., Nomoto, K., \&  Hashimoto, M. 1993, in Prantoz N., 
Vangoni-Flam, E., eds., Origin
and Evolution of the Elements, Cambridge Univ. Press, Cambridge, p.297

\bibitem{19}
Woosley, S.E. \& Weaver, T.A. 1995, ApJS, 87, 109

\end{thebibliography}
\end{document}